\documentclass[reprint,
amsmath,amssymb,aps,superscriptaddress%
]{revtex4-2}

\usepackage{graphicx}
\usepackage{dcolumn}
\usepackage{bm}
\usepackage[utf8]{inputenc}
\usepackage[T1]{fontenc}
\usepackage{amsmath}
\usepackage{hyperref}
\usepackage{subfig}
\usepackage{float}
\floatstyle{plaintop}
\restylefloat{table}

\begin{document}

\title{First-principles density limit scaling in tokamaks based on edge turbulent transport and implications for ITER}

\author{M. Giacomin}
 \email{maurizio.giacomin@epfl.ch}
\author{A. Pau}
\author{P. Ricci}
\author{O. Sauter}
\affiliation{Ecole Polytechnique Fédérale de Lausanne (EPFL), Swiss Plasma Center (SPC), CH-1015 Lausanne, Switzerland}
\author{T. Eich}
\affiliation{Max-Planck-Institute for Plasma Physics, Boltzmannstr. 2, D-85748 Garching, Germany}
\author{\\the ASDEX Upgrade team}
\affiliation{See Meyer \emph{et al} 2019 (\url{https://doi.org/10.1088/1741-4326/ab18b8}) for the ASDEX Upgrade team.}
\author{JET Contributors}
\affiliation{See the author list of ``Overview of JET results for optimising ITER operation'' by J. Mailloux \emph{et al} to be published in Nuclear Fusion Special issue: Overview and Summary Papers from the 28$^{th}$ Fusion Energy Conference (Nice, France, 10-15 May 2021)}
\author{the TCV team}
\affiliation{See Coda \emph{et al} 2019 (\url{https://doi.org/10.1088/1741-4326/ab25cb}) for the TCV team.}

\begin{abstract}

A first-principles scaling law, based on turbulent transport considerations, and a multi-machine database of density limit discharges from the ASDEX Upgrade, JET and TCV tokamaks, show that the increase of the boundary turbulent transport with the plasma collisionality sets the maximum density achievable in tokamaks. 
This scaling law shows a strong dependence on the heating power, therefore predicting for ITER a significantly larger safety margin than the Greenwald empirical scaling (Greenwald \emph{et al}, Nucl. Fusion, 28(12), 1988) in case of unintentional H-L transition. 

\end{abstract}

\maketitle

A burning plasma in tokamaks requires plasma densities near the operational limit in order to be economically viable. At the same time, it cannot tolerate any disruption.
A purely empirical scaling law for the density limit in tokamaks was obtained in 1988~\cite{greenwald1988},
\begin{equation}
\label{eqn:greenwald}
    n_{GW}[10^{20} \text{m}^{-3}] = \frac{I_p [\text{MA}]}{\pi a[\text{m}]^2},
\end{equation}
where $n_{GW}$, known as Greenwald density, is the predicted maximum line-averaged density, $I_p$ the plasma current and $a$ the plasma minor radius.
When exceeded, experimental investigations observe the onset of magnetohydrodynamics (MHD) modes leading to a plasma discharge disruption.

Experiments indicate that the cooling of tokamak boundary plasma together with the subsequent increase of the plasma edge collisionality is a strong factor limiting the maximum achievable density in tokamaks, providing a strong link between the edge collisionality and density limit~\cite{vershkov1974,fielding1977,greenwald2002,labombard2005,hong2017,schmid2017}.
A Multifaceted Asymmetric Radiation From the Edge (MARFE) is often observed when approaching the density limit~\cite{lipschultz1984,lipschultz1987,mertens1994,frigione1996,bernert2014h,maraschek2017}, supporting the hypothesis of a strong edge physics role.
Experiments also demonstrated that the Greenwald limit can be exceeded through pellet injection that mainly increases the core, and only weakly, the edge density~\cite{lang2020}.
Observations from TCV experiments show the density limit preceded by a collapse in the edge temperature followed by a temperature profile decrease in the core region, leading to changes in the $q$ profile and internal inductance~\cite{kirneva2014,sauter2014}. 
The change in internal inductance indicates a modified plasma current profile that is susceptible to tearing modes and, finally, plasma disruption~\cite{gates2012}.

In this Letter, we show the link between the collapse of the pressure gradient at the tokamak edge and the crossing of the L-mode disruptive density limit by leveraging first-principles theoretical considerations and a multi-machine database of density limit discharges of the ASDEX Upgrade (AUG), JET and TCV tokamaks. 
These discharges cover a wide range of values of the density, plasma current and heating power with several external heating systems, such as Neutral Beam Injection (NBI), Electron Cyclotron Resonance Heating (ECRH) and Ion Cyclotron Resonance Heating (ICRH), and different wall types (TCV has a carbon wall whereas the other machines feature metal walls).
In particular, the plasma current ranges from 0.1~MA to 2.5~MA, the toroidal magnetic field from 1.4~T to 3~T, the tokamak major radius from 0.9~m to 3~m and the power crossing the separatrix from 0.1~MW to 9~MW, yielding line-averaged electron density ranging between $2\times 10^{19}$~m$^{-3}$ and $1.1\times 10^{20}$~m$^{-3}$.
Thanks to the wide range of parameter values as well as the various heating systems and tokamak sizes, this database is particularly suitable to make reliable conclusions on ITER.  

\begin{figure*}
    \centering
    \includegraphics[scale=0.6]{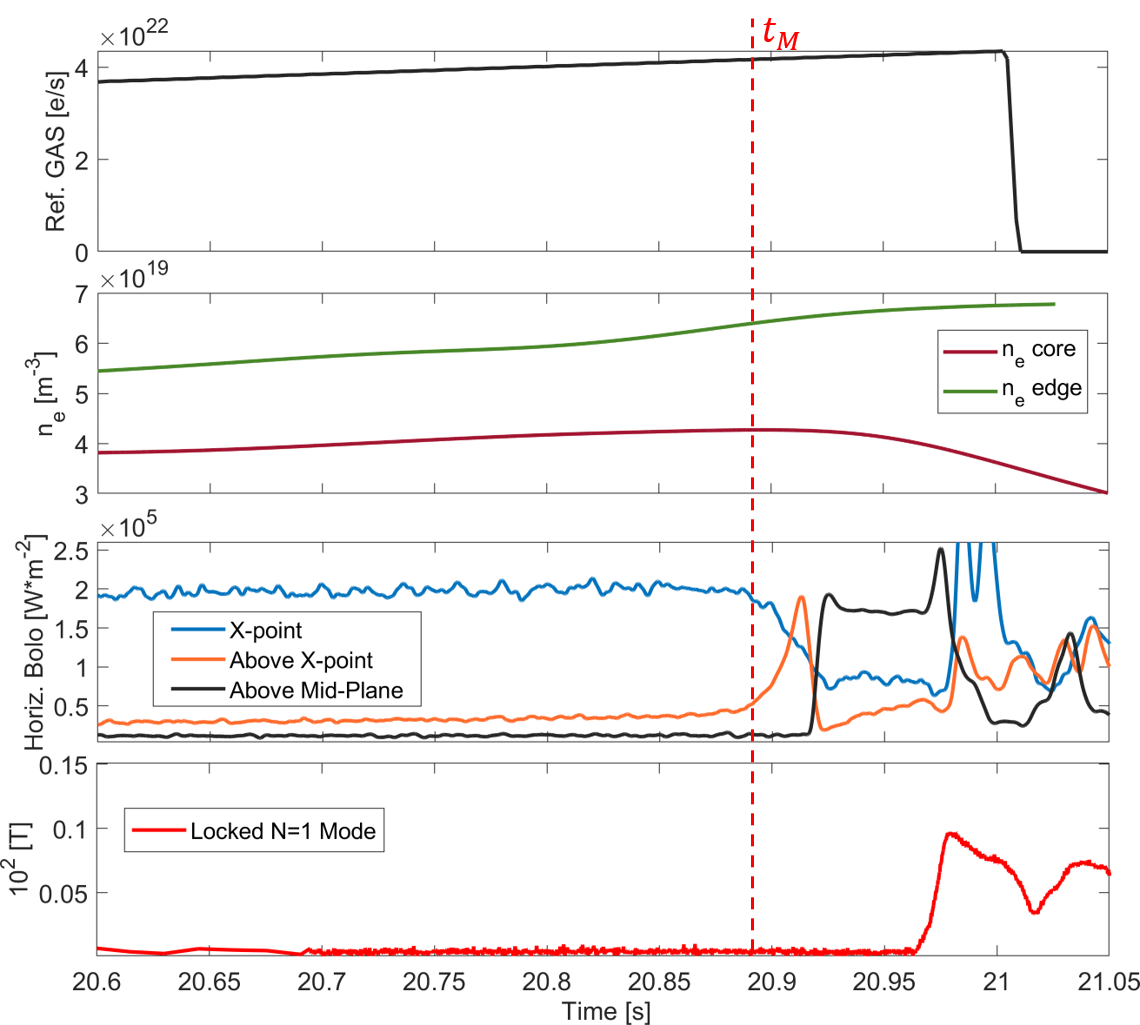}
    \caption{Time trace of the gas flux, electron density from Thomson scattering, radiation intensity and magnetic perturbations for the JET discharge \#80823. The MARFE event is identified by the strong increase of the radiation measured above the X-point. The MARFE onset precedes the appearance of a locked mode, which eventually leads to the plasma disruption. The red dashed vertical line represents the time of the MARFE onset, $t_M\simeq 20.9$~s. The onset of the locked $N=1$ mode occurs at 21.95~s, while the disruption time is at 21.1~s.}
    \label{fig:timetrace}
\end{figure*}

\begin{figure*}
    \centering
    \subfloat[]{\includegraphics[width=0.3\textwidth]{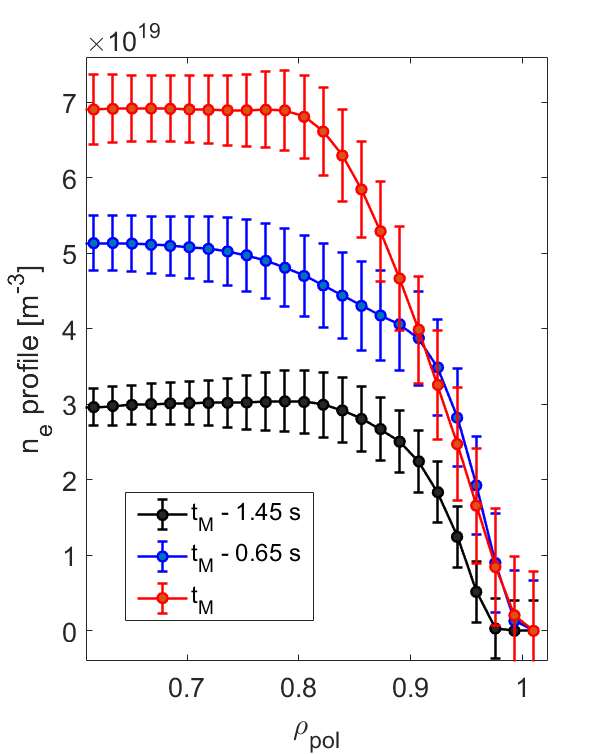}}\quad
    \subfloat[]{\includegraphics[width=0.3\textwidth]{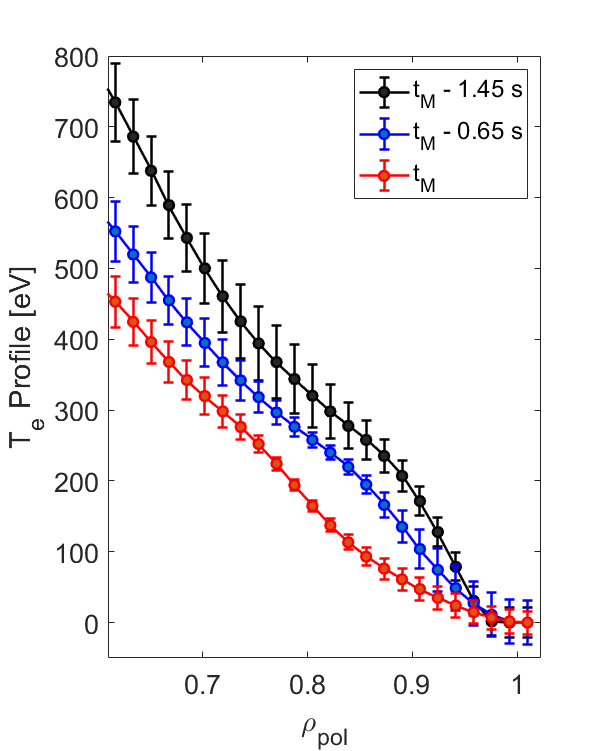}}\quad    \subfloat[]{\includegraphics[width=0.3\textwidth]{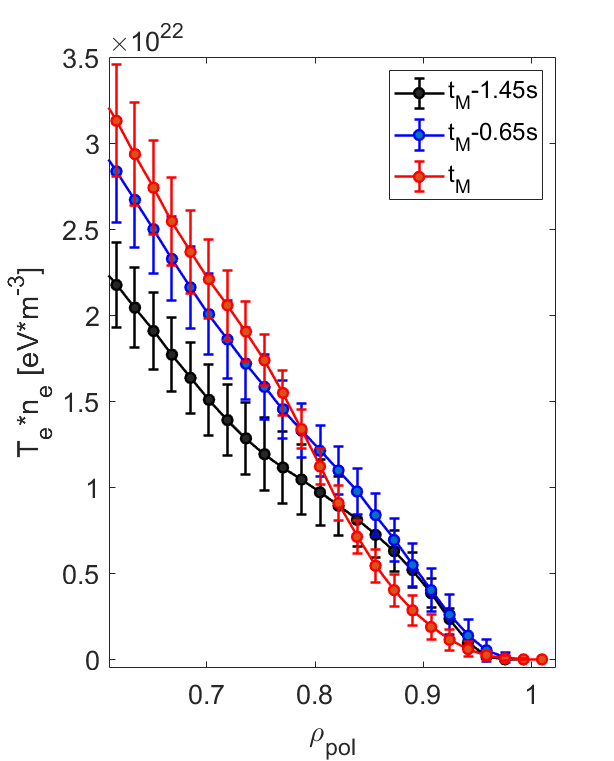}}
    \caption{Radial profile of the electron density (a), electron temperature (b) and electron pressure (c) as a function of the normalized radial coordinate at three different times of the JET discharge \#80823. The time of the MARFE onset is denoted as $t_M$ (see Fig.~\ref{fig:timetrace}). The radial profiles are obtained by fitting Thomson data with splines (Hermite form), imposing the profile and its first derivative to vanish at $\rho_\text{pol} = 1.1$ and enforcing the monotonicity of the profile in the edge region.  }
    \label{fig:profiles}
\end{figure*}

The density limit is often studied experimentally by increasing the gas flux, and thus plasma density, until causing an intentional disruption, as shown in Fig.~\ref{fig:timetrace} for the JET discharge \#80823~\cite{huber2013}. Here, the density increases until the onset of a MARFE, as identified by a strong increase in the radiation intensity in the region above the X-point (see Fig.~\ref{fig:timetrace}), that is followed by an MHD mode and plasma disruption. 
Fig.~\ref{fig:profiles} shows the radial profiles of the electron density, electron temperature and electron pressure at three times before the onset of the MARFE. 
The density increase is accompanied by a cooling of the electron temperature at the tokamak edge and an increase of the edge collisionality.
This plasma collisionality increase is expected to enhance turbulent transport, as experimentally shown in Ref.~\cite{labombard2001}, that further decreases the edge temperature and thus pressure gradients until they collapse (see Fig.~\ref{fig:profiles}).
Finally, the collapse of the edge temperature and pressure gradients, associated with the MARFE onset, leads to a reduced plasma current channel where the $q$ profile decrease triggers MHD modes.

As shown in Fig.~\ref{fig:profiles}, the crossing of the density limit can be associated with a collapse of the pressure gradient near the edge (between $\rho_\text{pol}=0.9$ and $\rho_\text{pol}=1$, where $\rho_\text{pol}=\sqrt{\psi/\psi_\text{LCFS}}$ with $\psi$ the poloidal flux and $\psi_\text{LCFS}$ the value of $\psi$ at the last-closed flux surface (LCFS)), which is estimated here from a balance between heat source, turbulent transport across the separatrix and parallel losses at the vessel wall.
The physical model considered here is based on a simplified drift-reduced Braginskii fluid model,
\begin{align}
\label{eqn:pressure}
\frac{\partial p_e}{\partial t} + \mathbf{v_E}\cdot\nabla p_e =  \nabla_\parallel(\chi_{\parallel e}\nabla_\parallel T_e)+s_p\,,\\
\label{eqn:vorticity}
\frac{\partial}{\partial t}\nabla_\perp^2 \phi =  \frac{\Omega_{ci} B}{e n}\biggl(\nabla\times\frac{\mathbf{b}}{B}\biggr)\cdot\nabla p_e- \frac{\Omega_{ci} B}{e n \sigma_\parallel}\nabla_\parallel^2\phi\,,
\end{align}
where $p_e = n T_e$ is the electron pressure, $n$ and $T_e$ the electron density and electron temperature, respectively, $\phi$ the electrostatic potential, $\mathbf{v_E}=(\mathbf{b}\times \nabla \phi)/B$ the $\mathbf{E}\times\mathbf{B}$ drift velocity, $\Omega_{ci}=e B/m_i$ the ion cyclotron frequency, $\nabla_\parallel f=\mathbf{b}\cdot\nabla f$ the gradient parallel to the magnetic field,  ${\nabla_\perp^2 f=\nabla\cdot\bigl[(\mathbf{b}\times\nabla f)\times\mathbf{b}\bigr]}$ the laplacian operator acting on the plane perpendicular to the magnetic field, with $f$ a scalar function, $B$ the modulus of the magnetic field, $\mathbf{b}=\mathbf{B}/B$ the unit vector of the magnetic field, $s_p$ the electron heat source, 
\begin{equation}
    \label{eqn:conduct}
    \chi_{\parallel e}=3.16\frac{2}{3}\frac{n T_e \tau_e}{m_e}
\end{equation}
the electron thermal parallel conductivity and
\begin{equation}
    \label{eqn:sigma}
    \sigma_\parallel = 1.96\frac{n e^2 \tau_e}{m_e}
\end{equation}
the parallel conductivity, with $\tau_e$ the electron collision time.
A fluid model approach is justified by the high plasma collisionality in the tokamak boundary for these conditions. At high collisionality, we expect boundary turbulence to be driven mainly by resistive interchange modes, as supported by nonlinear turbulent simulations~\cite{beadle2020,giacomin2020transp,giacomin2021,tatali2021}.
The effect of core radiation is included in the power entering into the scrape-off layer (SOL), $P_\text{SOL} = P_\text{tot} - P_\text{rad} \simeq 2\pi R_0 S_p = 2 \pi R_0 \int_{A_{\text{core}}} s_p \,\mathrm{d}A$, with $P_\text{tot}$ the total heating power, $P_\text{rad}$ the core radiated power and $A_\text{core}$ the area on a poloidal plane inside the LCFS.
We do not, conversely, consider the effect of  neutrals on the turbulent transport, as results from Ref.~\cite{mancini2021} indicate that neutral dynamics affects only weakly such turbulent transport near the separatrix. 
In the following, $p_e$ is written as the sum of an equilibrium component, $\bar{p}_e$, defined as the time and toroidal average of $p_e$, and its fluctuation, $\tilde{p}_e$, defined as $\tilde{p}_e=p_e-\bar{p}_e$ (and similarly for $\phi$).

In Ref.~\cite{rogers1998}, the density limit has been associated to a strong increase of turbulent transport due to nonlinear electromagnetic fluctuations, while Ref.~\cite{eich2021} associates it to a turbulent regime transition.
Here, we associate the crossing of the density limit to the collapse of the edge pressure gradient, which is estimated by the equilibrium pressure gradient length, $L_p = \bar{p}_e/|\partial_\psi \bar{p}_e|$, becoming a considerable fraction of the tokamak minor radius, i.e.  $L_p \sim a$.
An estimate of $L_p$ can be obtained by balancing $S_p$ with the cross-field heat flux $q_\psi$ integrated over the LCFS, i.e. 
\begin{equation}
    \label{eqn:balance_approx}
    S_p\sim \pi a\sqrt{\frac{1+\kappa^2}{2}} q_\psi\,,
\end{equation}
with $\kappa$ the plasma elongation.
In the highly turbulent transport regime considered here, the equilibrium cross-field heat flux across the separatrix can be neglected, i.e. $\bar{p}_e \bar{v}_{E,\psi} \ll \overline{\tilde{p}_e \tilde{v}_{E,\psi}}$, where $v_{E,\psi}$ is the component of $\mathbf{v}_{E}$ along $\nabla \psi$, with $\psi$ the flux function. The cross-field heat flux across the LCFS is therefore solely determined by turbulent transport, ${q_\psi \simeq \overline{\tilde{p}_e \tilde{v}_{E,\psi}}}$, where $\tilde{v}_{E,\psi}$ can be obtained by linearizing Eq.~\eqref{eqn:pressure}, $\gamma \tilde{p}_e \sim  - \partial_\psi \bar{p}_e \tilde{v}_{E,\psi}$, with $\gamma\simeq \sqrt{2/(R_0 L_p)} c_s$ the growth rate of the interchange instability~\cite{Mosetto2013}, $c_s=\sqrt{\bar{T}_e/m_i}$ the ion sound speed and $R_0$ the tokamak major radius. 
Following Ref.~\cite{giacomin2020transp}, $\tilde{p}_e$ is estimated by assuming the growth of the linearly unstable modes saturates when the instability drive is removed from the system~\cite{Ricci2013,Ricci2008}, i.e. $k_\psi \tilde{p}_e\sim \bar{p}_e/L_p$, where  $k_{\psi}\simeq\sqrt{k_{\chi}/L_p}$ as shown by non-local linear analysis~\cite{Ricci2008} and $k_{\chi} \simeq \Omega_{ci}/(q R_0) [m_i \sigma_\parallel/(e^2\bar{n}\gamma)]^{1/2}$  is obtained by balancing the interchange drive and the parallel current terms in Eq.~\eqref{eqn:vorticity}~\cite{Halpern2014}, where $q$ is the edge safety factor at the 95\% flux surface. 
These assumptions lead to
\begin{equation}
    \label{eqn:transport_int}
    q_\psi \sim 2^{3/4}\frac{c_s^{3/2} q R_0^{1/4}}{\Omega_{ci}L_p^{3/4}}\sqrt{\frac{e^2 \bar{n}}{m_i\sigma_\parallel}}\bar{p}_e\,.
\end{equation}
By substituting $q_\psi$, Eq.~\eqref{eqn:transport_int}, into Eq.~\eqref{eqn:balance_approx} and solving for $L_p$, we obtain
\begin{equation}
    \label{eqn:lp}
    L_p \sim \biggl[2\pi^4 a^4 (1+\kappa^2)^2 \frac{c_s^6 q^4 R_0}{\Omega_{ci}^4}  \biggl(\frac{e^2 \bar{n}}{m_i \sigma_\parallel}\biggr)^2 \biggl(\frac{ \bar{p}_e}{S_p}\biggr)^4\biggr]^{1/3}\,.
\end{equation}
We estimate $\bar{T}_e$ at the LCFS by balancing the heat source with the parallel heat losses in the SOL, by assuming parallel heat conduction dominates over parallel heat convection, as would be expected at high density.
This leads to~\cite{stangeby2000}
\begin{equation}
    \label{eqn:twopoint}
    \bar{T}_e \sim \biggl(\frac{7}{2}\frac{S_p L_\parallel}{\chi_{\parallel e0} L_p}\frac{q R_0}{a}\biggr)^{2/7}\,,
\end{equation}
where $\chi_{\parallel e0} = \chi_{\parallel e} \bar{T}_e^{-5/2}$, $L_\parallel$ is the parallel connection length in the SOL and we have approximated $B/B_\theta$ at the outboard midplane with $q R_0/a$ ($B_\theta$ is the modulus of the poloidal magnetic field).
A final scaling of $L_p$ is obtained by substituting $\bar{T}_e$, Eq.~\eqref{eqn:twopoint}, into Eq.~\eqref{eqn:lp},
\begin{widetext}
\begin{equation}
    \label{eqn:lp_final}
    L_p \sim  \frac{(1+\kappa^2)^{14/29}a^{20/29} R_0^{15/29} q^{36/29}\bar{n}^{28/29} L_\parallel^{8/29}}{m_i^{21/29}\Omega_{ci}^{28/29} S_p^{20/29}\chi_{\parallel e0}^{8/29}}\biggl(\frac{e^2\bar{n}}{m_i \sigma_{\parallel 0}}\biggr)^{14/29}\,,
\end{equation}
\end{widetext}
where $\sigma_{\parallel  0} = \sigma_\parallel \bar{T}_e^{-3/2}$ and we have dropped numerical constants. 

Imposing the condition $L_p\sim a$ in Eq.~\eqref{eqn:lp_final} leads to
\begin{equation}
\label{eqn:den_lim}
    n_\text{lim} \sim \frac{m_i^{5/6}\Omega_{ci}^{14/21}a^{3/14}S_p^{10/21}\chi_{\parallel e0}^{4/21}\sigma_{\parallel 0}^{1/3}}{e^{2/3}(1+\kappa^2)^{1/3}q^{6/7} R_0^{5/14} L_\parallel^{4/21}}\,,
\end{equation}
where $n_\text{lim}$ denotes the maximum density that can be achieved in the proximity of the separatrix. 
By taking $S_p = P_\text{SOL}/(2\pi R_0)$, $L_\parallel\simeq q R_0$ and using Eqs.~\eqref{eqn:conduct} and \eqref{eqn:sigma}, we can write this scaling for the maximum achievable edge density, Eq.~\eqref{eqn:den_lim}, in terms of engineering parameters, 
\begin{equation}
    \label{eqn:den_lim_fin}
    n_\text{lim} = \alpha A^{1/6} a^{3/14} P_\text{SOL}^{10/21} R_0^{-43/42} q^{-22/21}(1+\kappa^2)^{-1/3}B_T^{2/3}\,.
\end{equation}
with $n_\text{lim}$ in units of 10$^{20}$~m$^{-3}$, $A$ the mass number of the main plasma ions, $P_\text{SOL}$ the power crossing the separatrix in MW, $R_0$ and $a$ in m and $B_T$ the toroidal magnetic field in T.  
The parameter $\alpha$ is a numerical coefficient, of order unity, that accounts for all numerical constants and  approximations that remain from order of magnitude estimates.
This proportionality constant could depend upon the plasma shape and divertor geometry. 
However, as shown later, a unique value of $\alpha$ is sufficient to describe the maximum density achievable for all the tokamaks and discharges considered herein. 

To compare Eq.~\eqref{eqn:greenwald} and Eq.~\eqref{eqn:den_lim_fin}, we rewrite Eq.~\eqref{eqn:den_lim_fin} in terms of the plasma current, 
\begin{equation}
    \label{eqn:den_lim_ip}
     n_\text{lim} \sim  A^{1/6} P_\text{SOL}^{10/21}R_0^{1/42}B_T^{-8/21}(1+\kappa^2)^{-1/3}\frac{I_p^{22/21}}{a^{79/42}}\,.
\end{equation}
We note that Eq.~\eqref{eqn:greenwald} and Eq.~\eqref{eqn:den_lim_ip} share a main dependence on $I_p$ and $a$, but the density limit in Eq.~\eqref{eqn:den_lim_ip} now depends on $P_\text{SOL}$. 
The power dependence in the density limit has been extensively investigated in the past (see e.g. Refs.~\cite{stabler1992,mertens1997,rapp1999,mertens2000,borrass2004,esposito2008,huber2013,bernert2014h}), leading to contrasting results, where either significant or no dependence on the heating power has been observed.
Focusing on the L-mode disruptive density limit considered here, while no power dependence is reported in early references~\cite{greenwald1988,greenwald2002}, a more recent experimental investigation based on JET data has found a significant dependence on the heating power, with the maximum achievable density proportional to $P_\text{heat}^{0.4}$~\cite{huber2013}, which agrees well with the power dependence in theoretical scaling of Eq.~\eqref{eqn:den_lim_ip}, i.e. $n_\text{lim} \propto P_\text{SOL}^{0.48}$.
We also highlight that a significant dependence on the heating power is also found in a recent density limit scaling derived in  Refs.~\cite{zanca2017,zanca2019} from a thermal balance between the heating power, the radiative emission and the radial transport based upon an effective perpendicular heat diffusivity.

We now validate Eq.~\eqref{eqn:den_lim_fin} against the multi-machine database of density limit discharges from the AUG, JET and TCV tokamaks. Two different scenarios are considered: (i) a standard scenario where the density limit is reached in L-mode (i.e. no H-mode phase), where the plasma density is increased up to the density limit, and (ii) an ITER-relevant scenario where the L-mode density limit is preceded by an H-mode phase~\cite{huber2013,bernert2014h,maraschek2017,lang2020,vu2021}. 
In the second scenario, the plasma undergoes first an L-H transition, then as the density is increased, plasma confinement degrades until an H-L transition occurs and, once in L-mode, a density limit is attained. 

\begin{figure*}[t!]
    \centering
    \subfloat[Theoretical scaling]{\includegraphics[height=0.25\textheight]{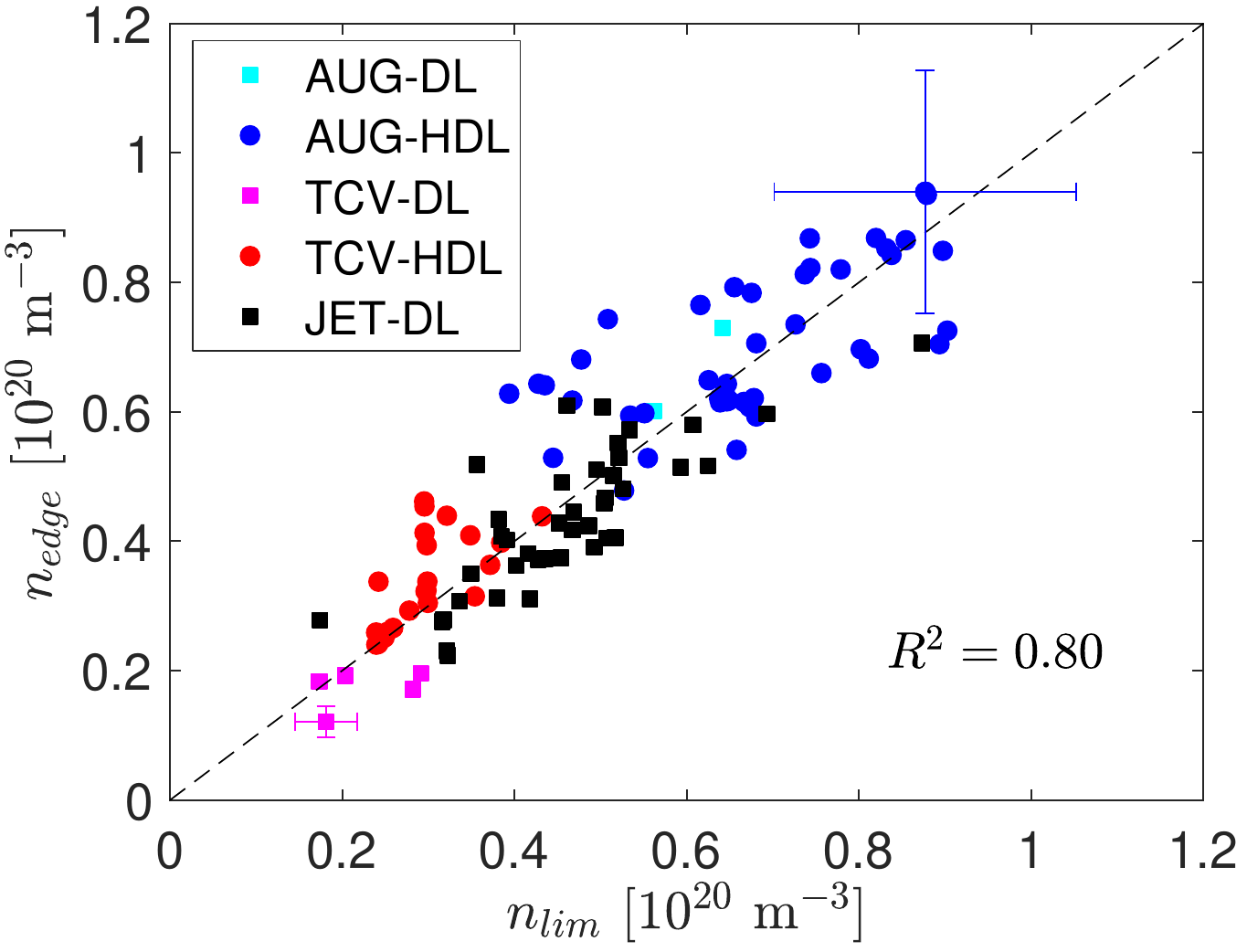}}\quad
    \subfloat[Empirical scaling]{\includegraphics[height=0.25\textheight]{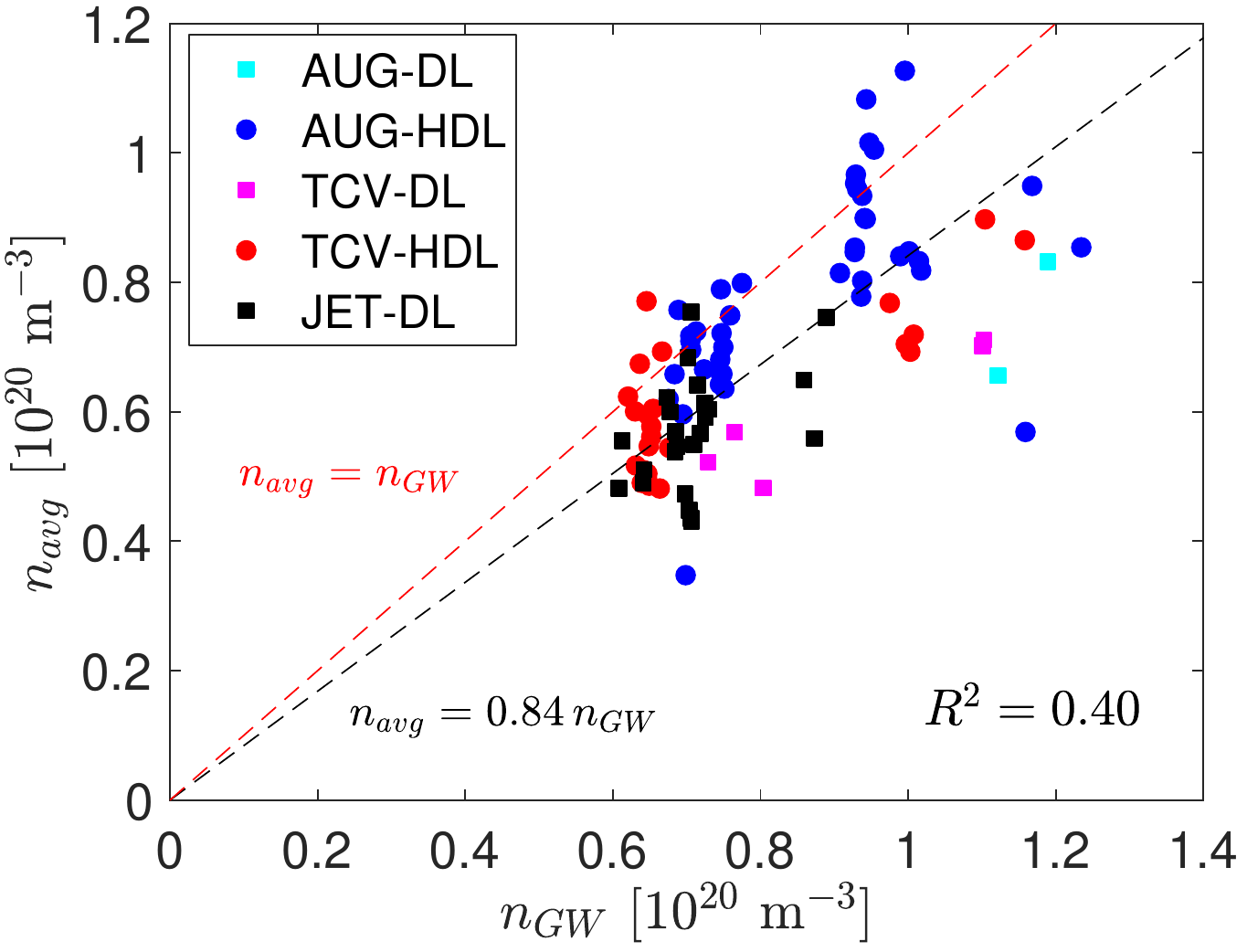}}
    \caption{(a) Experimental edge density measured by means of Thomson scattering at the onset of the MARFE compared to the theoretical prediction $n_\text{lim}$ provided by Eq.~\eqref{eqn:den_lim_fin} with $\alpha=3.3\pm 0.3$. (b) Experimental measured maximum line-averaged density ($n_\text{avg}$) compared to the prediction provided by the empirical scaling law in Eq.~\eqref{eqn:greenwald} (red dashed line). For comparison purpose, a proportionality factor is also considered in the empirical scaling when evaluating the $R^2$ parameter. 
    Different marker shapes and colors are used to distinguish between the standard density limit (DL) and the density limit of discharges with an H-mode phase (HDL). As examples, the errorbars are shown for the discharge with the lowest and highest density.}
    \label{fig:comparison}
\end{figure*}

Validation is performed by fitting the experimental edge density at the MARFE onset to the  scaling in Eq.~\eqref{eqn:den_lim_fin}, using the numerical factor $\alpha$ as the only fitting parameter that is the same in all discharges on all the tokamaks.
The choice of the MARFE onset as the reference time is motivated by considering the MARFE as a precursor of the density limit.
The experimental value of the density at the MARFE onset is obtained by averaging the edge density, measured by means of Thomson scattering, in the region between $\rho_\text{pol}=0.85$ and $\rho_\text{pol}=0.95$ in all the discharges and tokamaks.
The radial interval between $\rho_\text{pol}=0.85$ and $\rho_\text{pol}=0.95$ is chosen to reduce uncertainties of the experimental density. While choosing the interval between $\rho_\text{pol}=0.9$ and $\rho_\text{pol}=1.0$ does not affect the overall trend, it adds significant uncertainties on the experimental values of density.

The result of the comparison is shown in Fig.~\ref{fig:comparison}~(a).
The theoretical scaling law well reproduces the measured edge density at MARFE onset, with a high fitting quality parameter, $R^2\simeq 0.8$,
and $\alpha \simeq 3.3\pm 0.3$. Fitting separately the three tokamaks leads to variances in $\alpha$ that are below 10~\%, showing the robustness of this approach with respect to machine specificities. 
We underline that L-mode and H-mode scenarios both follow our scaling.
The density limit for both scenarios is thus described with the same plasma dynamics, independently of discharge history and/or wall type.

The uncertainty on the theoretical predictions is mainly due to the experimental measurement of the power crossing the separatrix, which is estimated from the total power coupled to the plasma, having subtracting the core radiated power. The later is inferred by line integrated measurements of the bolometer cameras looking at the main plasma (typically by a tomographic inversion) and is affected by an experimental uncertainty that can be up to 50~\%.
In order to reduce the experimental uncertainty, the power crossing the separatrix is averaged on a time window of 20~ms before the MARFE onset, excluding the values close to the MARFE event where this quantity drops significantly because of the strong increase of the core radiated power.
As indicative value, we estimate a 20~\% uncertainty for the experimental values of the edge density as well as for its theoretical prediction.

Fig.~\ref{fig:comparison}~(b) shows a comparison between the maximum line-averaged density of the discharges considered here to the empirical scaling law predictions from Eq.~\eqref{eqn:greenwald} (red line in Fig.~\ref{fig:comparison}~(b)). 
The quality of the agreement is evaluated through the parameter $R^2$ and a proportionality constant is also used for the empirical scaling (black line in Fig.~\ref{fig:comparison}~(b)). 
Although the empirical scaling is able to reproduce the overall experimental trend, there is a significant scatter in experimental data indicating missing dependencies in Eq.~\eqref{eqn:greenwald} and low $R^2$.
A dependence on $B_T$ appears in Eq.~\eqref{eqn:den_lim_ip}, which is not present in Eq.~\eqref{eqn:greenwald}. This dependence leads to a decrease of the density limit in high-$B_T$ fusion devices. For example, Eq.~\eqref{eqn:den_lim_fin} yields to $n_\text{lim}=5\times 10^{20}$~m$^{-3}$ for Alcator C-Mod ($R_0 = 0.67$~m, $a = 0.22$~m, $\kappa=1.5$, $B_T = 8$~T, $P_\text{SOL} =5$~MW and $q=4$~\cite{greenwald2014}), which agrees with the maximum density achieved in C-Mod shown in Fig.~3 of Ref.~\cite{greenwald2002}, and it is approximately a factor of two smaller than the prediction of Eq.~\eqref{eqn:greenwald}.



We conclude with a prediction of the density limit for SPARC~\cite{creely2020} ($R_0=1.85$~m, $a=0.57$~m, $B_T=12.2$~T, $q=3$, $\kappa =2$) and ITER ($R_0=6.2$~m, $a=2$~m, $B_T=5.3$~T, $q=3$ and $\kappa=1.8$).
The scaling in Eq.~\eqref{eqn:den_lim_fin} with SPARC parameters and $P_\text{SOL} \simeq 28$~MW leads to $n_\text{lim}^\text{\tiny{SPARC}} \simeq 8.7\times 10^{20}$~m$^{-3}$, which is very close to the Greenwald density for SPARC, $n_{GW}^\text{\tiny{SPARC}} \simeq 8.5\times 10^{20}$~m$^{-3}$.
Although SPARC will probably not provide a definitive data point that would decide between Eq.~\eqref{eqn:greenwald} and Eq.~\eqref{eqn:den_lim_fin}, it will operate well below both limits~\cite{creely2020}.
On the other hand, the scaling in Eq.~\eqref{eqn:den_lim_fin} with ITER parameters and $P_\text{SOL} \simeq 50$~MW leads to $n_\text{lim}^\text{\tiny{ITER}} \simeq 2.5\times 10^{20}$~m$^{-3}$, which is a factor of two higher than the Greenwald density for ITER, $n_{GW}^\text{\tiny{ITER}} \simeq 1.2\times 10^{20}$~m$^{-3}$. 
We underline that fusion power plants will operate with a much larger $P_\text{SOL}$ than present day tokamaks, leading to significantly higher values of density limit than the Greenwald scaling and therefore a larger safety margin in case of accidental transition to L-mode, with important implications for the design and operation of future fusion power plants.
We stress that our predictions rely on purely physics-based calculations and on a database of discharges spanning a larger size than the distance to ITER. 
Nevertheless, given the important consequences for ITER, this result calls for the urgent need of further experimental investigations of the power dependence in the L-mode density limit.  

This work was carried out within the framework of the EUROfusion Consortium and has received funding from the Euratom research and training programme 2014 - 2018 and 2019 - 2020 under grant agreement No 633053. The views and opinions expressed herein do not necessarily reflect those of the European Commission. 
This work was supported in part by the Swiss National Science Foundation.



\bibliographystyle{unsrt}
\bibliography{bibliography}

\end{document}